\documentclass[review, 3p]{elsarticle}

\usepackage{lineno}

\usepackage{graphicx,psfrag,epsf}
\usepackage{subfigure}
\RequirePackage[colorlinks,citecolor=blue,urlcolor=blue]{hyperref}

\usepackage{amsmath,amssymb}
\usepackage{amsthm}
\usepackage{times}
\usepackage{bm}
\usepackage{natbib}
\usepackage{amssymb}
\usepackage{setspace}
\usepackage{bbm}
\usepackage{tikz}
\usetikzlibrary{positioning,shapes,shadows,arrows}
\usepackage{scrextend}
\usepackage{cleveref}% l
\usepackage{smartdiagram}
\usepackage[all]{nowidow}

\biboptions{authoryear}

\usepackage{tikz}
\usetikzlibrary{fit,positioning,calc}
\tikzset{
  font={\fontsize{11pt}{12}\selectfont}}
\usepackage[ruled,norelsize,vlined]{algorithm2e}

\newtheorem{remark}{Remark}
\newtheorem{theorem}{Theorem}

\newtheorem{proposition}{Proposition}

\begin{document}

\begin{frontmatter}

\title{Tractable Bayesian density regression via logit stick-breaking priors}

\author[a]{Tommaso Rigon\corref{mycorrespondingauthor}}
\cortext[mycorrespondingauthor]{Corresponding author}
\ead{tommaso.rigon@duke.edu}
\address[a]{Department of Statistical Science, Duke University, Durham, NC, USA}

\author[b]{Daniele Durante}
\ead{daniele.durante@unibocconi.it}
\address[b]{Department of Decision Sciences and Bocconi Institute for Data Science and Analytics, Bocconi University, Via Roentgen 1, 20136 Milano, Italy}

\begin{abstract}
There is a growing interest in learning how the distribution of a response variable changes with a set of predictors.  Bayesian nonparametric dependent mixture models provide a flexible approach to address this goal. However, several formulations require computationally demanding algorithms for posterior inference. Motivated by this issue, we study a class of predictor-dependent infinite mixture models, which relies on a simple representation of the stick-breaking prior via sequential logistic regressions. This formulation maintains the same desirable properties of popular predictor-dependent stick-breaking priors, and leverages a recent P\'olya-gamma data augmentation to facilitate the implementation of several computational methods for posterior inference. These routines include Markov chain Monte Carlo via Gibbs sampling, expectation-maximization algorithms, and mean-field variational Bayes for scalable inference, thereby stimulating a wider implementation of Bayesian density regression by  practitioners. The algorithms associated with these methods are presented in detail and tested in a toxicology study.
\end{abstract}


\end{frontmatter}

%\linenumbers

%##########################################################
\section{Introduction}
\label{sec1}
There is a growing interest in density regression models which allow the entire distribution of a univariate response variable $y \in \mathcal{Y}$ to be unknown and changing with a vector of predictors $\bf{x}  \in \mathcal{X}$. Indeed, the increased flexibility provided by these procedures allows improvements in inference and prediction compared to classical regression frameworks, as seen in  applications \citep[e.g.][]{DP08, GRI11, WDPT14}.

Within the Bayesian nonparametric framework, there is a wide variety of alternative methods to provide flexible inference for conditional distributions. Most of these strategies represent generalizations of the marginal density estimation problem for $f(y)$,  which is commonly addressed via Bayesian nonparametric mixture models of the form $f(y)= \int K(y ; \boldsymbol{\theta}) p(\mbox{d} \boldsymbol{\theta})$, where $ K(y ; \boldsymbol{\theta})$ is a known parametric kernel indexed by $\boldsymbol{\theta} \in \boldsymbol{\Theta}$, and $p$ is a random probability measure which is assigned a flexible prior $\Pi$. Popular choices for $\Pi$ are the Dirichlet process \citep{FE73}, the two-parameter Poisson-Dirichlet process \citep{PY97}, and other random measures having a stick-breaking representation \citep{IJ01}. This choice leads to the infinite mixture model  
\begin{eqnarray}
f(y)= \int K(y ; \boldsymbol{\theta})  p( \mbox{d} \boldsymbol{\theta})=\sum_{h=1}^{\infty} \pi_h K(y ; \boldsymbol{\theta}_h), 
\label{mix_intro}
\end{eqnarray}
where $\pi_h=\nu_h \prod_{l=1}^{h-1}(1-\nu_l)$ for every $h\geq 1$, with $\pi_1 = \nu_1$. In equation~\eqref{mix_intro}, the kernel parameters  $(\boldsymbol{\theta}_h)_{h\ge 1}$ are distributed according to a diffuse base measure  $P_0$, whereas the stick-breaking weights $(\nu_h)_{h\ge1}$ have independent $\mbox{Beta}(a_h,b_h)$ priors, so that $\sum_{h=1}^{\infty}\pi_h=1$ almost surely.  

Model \eqref{mix_intro} has key computational benefits in allowing the implementation of simple Markov chain Monte Carlo methods for posterior inference \citep[e.g.][]{EW95, NEA00}, and provides a consistent strategy for density estimation \citep[e.g.][]{GHO99, TOK06, GHO07}.  This has motivated different generalizations of \eqref{mix_intro} to incorporate the conditional density inference problem for $f(y \mid {\bf{x}})=f_{{\bf x}}(y)$, by allowing the random mixing measure $p_{\bf{x} }$ to change with  $\bf{x}  \in \mathcal{X}$, under a dependent stick-breaking characterization \citep{MC99,MC00}. Popular representations consider predictor-independent mixing weights $\pi_h$, and incorporate changes with $\bf{x}  \in \mathcal{X}$ in the atoms $\boldsymbol{\theta}_h(\bf{x} )$; see for instance \citet{DEI04,GEL05,CRU07}. As noted in \citet{MC00} and \citet{GS06}, the predictor-independent assumption for the mixing weights might have limited flexibility in practice. This has motivated more general formulations allowing also  $\pi_h(\bf{x} )$ to change with the predictors. Relevant examples include the order-based dependent Dirichlet process \citep{GS06}, the kernel stick-breaking process \citep{DP08}, the infinite mixture model with predictor-dependent weights  \citep{AWW14}, and more recent representations for Bayesian dynamic inference \citep{GUT16}. These formulations provide a broader class of priors for Bayesian density regression, but their flexibility comes at a computational cost. In particular, the availability of simple algorithms for tractable posterior inference is limited by the specific construction of these representations. 

The above issues motivate alternative formulations which preserve theoretical properties, but facilitate tractable posterior computation under a broader variety of  algorithms. We aim to address this goal via a logit stick-breaking prior (\textsc{lsbp}), which relates each stick-breaking weight $\nu_h({\bf{x}} ) \in (0,1)$ to a function  $\eta_h(\bf{x} ) \in \Re$ of the covariates,  using the logit link. The proposed formulation is closely related to the probit stick-breaking prior (\textsc{psbp}) of \citet{RD11}. Indeed, as we will discuss in Section~\ref{Section2}, both \textsc{lsbp} and  \textsc{psbp} are characterized by a continuation-ratio representation \citep{TU91}, which allows to express the underlying clustering assignment in terms of independent and sequential binary regressions. This representation has key computational benefits and has been exploited by~\citet{RD11} to derive a Markov chain Monte Carlo (\textsc{mcmc}) algorithm for posterior inference. However, while the \textsc{mcmc} for \textsc{psbp} relies on the truncated Gaussian data augmentation for probit regression \citep{AC93}, the one for \textsc{lsbp} exploits the recent P\'olya-gamma data augmentation for logistic regression \citep{PS13}, which might improve mixing compared to the \textsc{psbp}, especially in  imbalanced situations \citep{JO18}. As we will clarify in Section~\ref{Section2}, these imbalanced settings can also occur in our case, since the binary regressions are associated to latent clustering allocations. We illustrate the \textsc{mcmc} algorithm for the \textsc{lsbp} in Section~\ref{Section3}.

Besides developing tractable Gibbs sampling methods, we further derive alternative computational routines which address the scalability and mixing issues of  \textsc{mcmc} in high-dimensional studies. Specifically, in Section~\ref{Section3} we illustrate a tractable expectation-maximization (\textsc{em}) routine for point estimation, and a simple variational Bayes (\textsc{vb}) algorithm for scalable inference. Both strategies leverage again the sequential representation of the \textsc{lsbp} and the associated P\'olya-gamma data augmentation. Note that a \textsc{vb} routine for  \textsc{lsbp} is also presented in \citet{RDC11}, but it is based on the bound of \citet{JJ00}. As a consequence of the recent theoretical findings in \citet{Durante2019}, it can be shown that our approach is intimately related to the one of \citet{RDC11}, although being developed by means of seemingly unrelated strategies. Finally, while tractable algorithms such as \textsc{em} or \textsc{vb} could be possibly obtained also for \textsc{psbp}, we are not aware of any actual discussion or implementation. Indeed, the analytical derivations might be slightly more complex in the \textsc{psbp} case compared to the \textsc{lsbp}, as discussed in Section~\ref{Section3}.

We shall emphasize that the overarching focus of our contribution is not on developing a novel methodological framework for Bayesian density regression, but on deriving a broad set of routine-use computational strategies under a suitable and tractable representation. To our knowledge this goal remains partially unaddressed, but represents a fundamental condition to facilitate routine implementation of Bayesian density regression by practitioners.  The three proposed algorithms are empirically compared in Section~\ref{Section4} using a real data toxicology study, previously considered in \citet{DP08}.  Section~\ref{Section5} provides concluding remarks.

\section{Logit stick-breaking prior}
\label{Section2}
This section presents a formal construction of the  \textsc{lsbp} via continuation-ratio logistic regressions. As a natural extension of model~\eqref{mix_intro}, we consider the general class of predictor-dependent infinite mixture models
\begin{eqnarray}
f_{\bf x}(y) =\int K_{{\bf x}}(y ; \boldsymbol{\theta})  p_{\bf{x}}(\mbox{d}\boldsymbol{\theta})   =\sum_{h=1}^{\infty} \pi_h({\bf x}) K_{\bf x}(y ; \boldsymbol{\theta}_h), 
\label{mix_dep}
\end{eqnarray}
where $\pi_h({\bf x})=\nu_h({\bf x}) \prod_{l=1}^{h-1}\{1-\nu_l({\bf x})\}$ are predictor-dependent mixing probabilities having a stick-breaking representation, whereas $K_{{\bf x}}(y ; \boldsymbol{\theta})$ denotes a predictor-dependent kernel, indexed by parameters $\boldsymbol{\theta}$ and covariates ${\bf x}$. 

To highlight the continuation-ratio representation of the  \textsc{lsbp}, let us first consider an equivalent formulation of the predictor-dependent mixture model in \eqref{mix_dep}. In particular, following standard hierarchical representations of mixture models, independent samples $y_1, \ldots, y_n$ of the variable with density function displayed in \eqref{mix_dep}, can be obtained from
\begin{eqnarray}
(y_i \mid  G_i= h, {\bf x}_i )\sim K_{{\bf x}_i}(y_i ; \boldsymbol{\theta}_h), \quad \mbox{with} \ \
  \mbox{pr}(G_i=h \mid {\bf x}_i)=\pi_h({\bf x}_i)=\nu_h({\bf x}_i) \prod_{l=1}^{h-1}\{1-\nu_l({\bf x}_i)\}, 
\label{marg}
\end{eqnarray}
for each unit $i=1, \ldots, n$, where $\boldsymbol{\theta}_h \sim P_0$ independently for $h \in \mathbb{N}$, whereas $G_i \in \mathbb{N}$ is the categorical variable denoting the mixture component associated with the $i$th unit. According to \eqref{marg}, every $G_i$ has probability mass function  $f(G_i \mid {\bf x}_i)=\prod_{h=1}^{\infty}\pi_{h}({\bf x}_i)^{\mathbbm{1}(G_i=h)}$, where $\mathbbm{1}(\cdot)$ denotes the indicator function. Hence, re-writing  $\{\nu_h({\bf x}_i)\}_{h\in \mathbb{N}}$ as a function of the mixing probabilities $\{\pi_h({\bf x}_i)\}_{h\in \mathbb{N}}$ via 
\begin{eqnarray}
 \nu_h({\bf x}_i)=\frac{\pi_h({\bf x}_i)}{1-\sum_{l=1}^{h-1}\pi_l({\bf x}_i)}=\frac{\mbox{pr}(G_i=h \mid {\bf x}_i)}{\mbox{pr}(G_i>h-1 \mid {\bf x}_i)}, \qquad h \in \mathbb{N},
 \label{cont_ratio}
\end{eqnarray}
allows to interpret each $\nu_h({\bf x}_i)$  as the probability of being allocated to component $h$, conditionally on the event of surviving to the previous $1, \ldots, h-1$ components, namely $\nu_h({\bf x}_i)=\mbox{pr}(G_i=h \mid G_i>h-1, {\bf x}_i)$. This result provides a formal characterization of the stick-breaking construction in \eqref{marg} as the continuation-ratio parameterization \citep{TU91} of the probability mass function for each component membership variable $G_i$. This connection with the literature on sequential inference for categorical data is common to all the stick-breaking priors---as mentioned also by \citet{RD11} in the probit case. 

\tikzstyle{abstract}=[rectangle, draw=black, rounded corners, anchor=north, text=black, text width=6cm]
\tikzstyle{myarrow}=[<-, >=open triangle 90, thick]
\tikzstyle{line}=[-, thick]
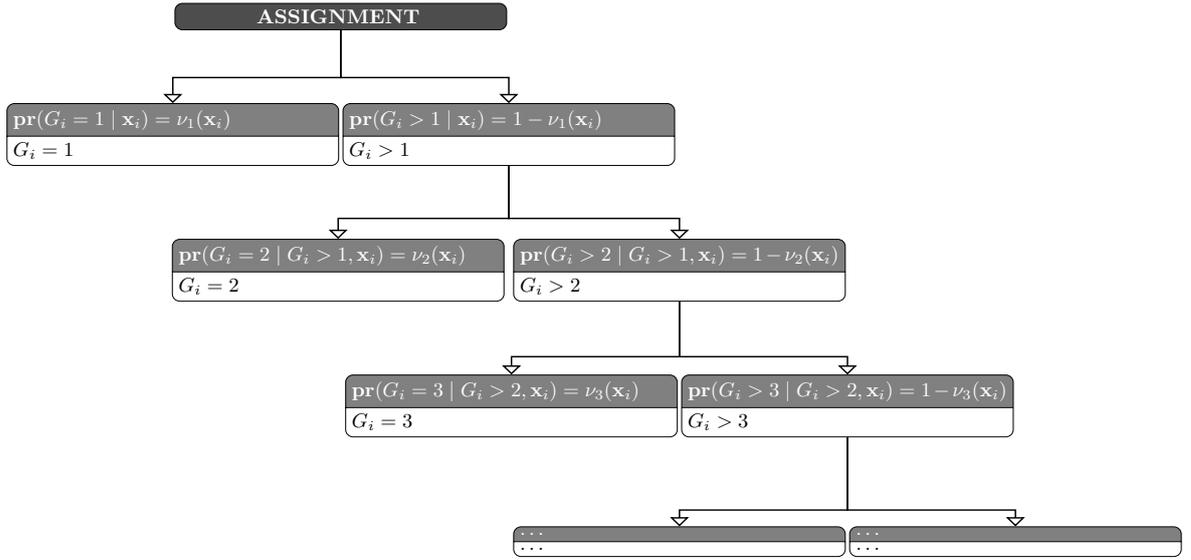
\begin{figure}[t]        
\begin{center}
\resizebox{16cm}{!}{
\begin{tikzpicture}[node distance=1.91cm]
    \node(Item) [abstract, rectangle split, rectangle split parts=1,minimum size=5pt,rectangle split part fill={black!70,white},]
        {
            \textcolor{white}{ \ \ \ \ \ \ \ \ \ \ \ \textbf{ASSIGNMENT}}
               };
    \node (AuxNode01) [text width=4cm, below=of Item] {};
    \node (Component) [abstract, rectangle split, rectangle split parts=2,rectangle split part fill={gray,white}, right=of AuxNode01,xshift=-4cm]
        {
            \textcolor{white}{{$\mbox{pr}(G_i >1 \mid {\bf{x}}_i )=1-\nu_1({\bf x}_i)$}}
            \nodepart{second} \texttt{$G_i >1$}
        };
    \node (System) [abstract, rectangle split, rectangle split parts=2,rectangle split part fill={gray,white}, left=of AuxNode01,xshift=4cm]
        {
            \textcolor{white}{{$\mbox{pr}(G_i=1 \mid {\bf{x}}_i )=\nu_1({\bf x}_i)$}}
            \nodepart{second} \texttt{$G_i=1$}
        };
    \node (AuxNode02) [text width=0.5cm, below=of Component] {};
    \node (Sensor) [abstract, rectangle split, rectangle split parts=2,rectangle split part fill={gray,white}, right=of AuxNode02,xshift=-2.2cm]
        {
            \textcolor{white}{{$\mbox{pr}(G_i>2 \mid  G_i>1,{\bf{x}}_i  )=1- \nu_2({\bf x}_i)$}}
            \nodepart{second}  \texttt{$G_i >2 $}
        };
    \node (Part) [abstract, rectangle split, rectangle split parts=2,rectangle split part fill={gray,white}, left=of AuxNode02,xshift=2.2cm]
        {
            \textcolor{white}{{$\mbox{pr}(G_i=2 \mid  G_i>1,{\bf{x}}_i  )=\nu_2({\bf x}_i)$}}
            \nodepart{second}  \texttt{$G_i =2 $}
        };
        
    \node (AuxNode03) [below=of Sensor] {};
    \node (Pressure) [abstract, rectangle split, rectangle split parts=2,rectangle split part fill={gray,white}, right=of AuxNode03, xshift=-2cm]
        {
            \textcolor{white}{{$\mbox{pr}(G_i>3 \mid  G_i>2,{\bf{x}}_i  )=1-\nu_3({\bf x}_i)$}}
            \nodepart{second}   \texttt{$G_i >3 $}
        };
    \node (Temperature) [abstract, rectangle split, rectangle split parts=2,rectangle split part fill={gray,white}, left=of AuxNode03, xshift=2cm]
        {
            \textcolor{white}{{$\mbox{pr}(G_i=3 \mid  G_i>2,{\bf{x}}_i  )=\nu_3({\bf x}_i)$}}
            \nodepart{second}  \texttt{$G_i =3 $} 
        };
   \node (AuxNode04) [below=of Pressure] {};
    \node (Pressure1) [abstract, rectangle split, rectangle split parts=2,rectangle split part fill={gray,white}, right=of AuxNode04, xshift=-2cm]
        {
            \textcolor{white}{{$\ldots$}}
            \nodepart{second}\texttt{$\ldots$}
        };
    \node (Temperature1) [abstract, rectangle split, rectangle split parts=2,rectangle split part fill={gray,white}, left=of AuxNode04, xshift=2cm]
        {
            \textcolor{white}{{$\ldots$}}
            \nodepart{second}  \texttt{$\ldots$}
        };
  
       \draw[myarrow] (Component.north) -- ++ (0,0.5) -| (Item.south);
        \draw[myarrow] (System.north) -- +(0,0.5) -| (Item.south);
        
          \draw[myarrow] (Sensor.north) -- ++ (0,0.4) -| (Component.south);
        \draw[myarrow] (Part.north) -- +(0,0.4) -| (Component.south);
        
          \draw[myarrow] (Pressure.north) -- ++ (0,0.35) -| (Sensor.south);
        \draw[myarrow] (Temperature.north) -- +(0,0.35) -| (Sensor.south);
        
          \draw[myarrow] (Pressure1.north) -- ++ (0,0.32) -| (Pressure.south);
        \draw[myarrow] (Temperature1.north) -- +(0,0.32) -| (Pressure.south);
   \end{tikzpicture}}
\caption{Representation of the sequential mechanism to sample $G_i$.}\label{F_decision}
\end{center}
\vspace{-5pt}
\end{figure}

As we will describe in Section~\ref{Section3}, the above result facilitates the implementation of  different routine-use algorithms in Bayesian inference, and provides a simple generative process for each $G_i$. In particular, as illustrated in Figure \ref{F_decision}, in the first step of this continuation-ratio generative mechanism, unit $i$ is either assigned to the first component  with probability $\nu_1({\bf{x}}_i)$ or to one of the others with complement probability. If $G_i=1$ the process stops, otherwise  it continues considering the reduced set $\{h: h>1\}$. A generic step $h$ is reached if $i$ has not been assigned to $1, \ldots, h-1$, and the decision at this step will be to either allocate $i$ to component $h$ with probability $\nu_h({\bf{x}}_i)$, or to one of the subsequent components  with probability $1-\nu_h({\bf{x}}_i)$, conditioned on $G_i >h-1$. Based on this representation, the assignment indicator $\zeta_{ih} = \mathbbm{1}(G_i=h)$ can be expressed, for every unit $i=1, \ldots, n$, as
\begin{eqnarray}
\zeta_{ih}=z_{ih} \prod_{l=1}^{h-1}(1-z_{il}),  \qquad h \in \mathbb{N},  
\label{bin_seq}
\end{eqnarray}
where the generic $z_{ih}$, $h \in \mathbb{N}$, is  a Bernoulli variable $(z_{ih} \mid {\bf{x}}_i) \sim \mbox{Bern}\{\nu_h({\bf{x}}_i)\}$ denoting the decision at the $h$th step to either allocate $i$ to component $h$ or to one of the subsequents. Hence, according to \eqref{bin_seq}, the  sampling of each $G_i$, under the predictor-dependent stick-breaking representation for each $\pi_h({\bf{x}}_i)$ in \eqref{marg}, can be reformulated as a set of sequential Bernoulli choices  with natural parameters $\eta_h({\bf x}_i )=\mbox{logit}\{\nu_h({\bf{x}}_i )\}=\log[\nu_h({\bf x}_i) /\{1-\nu_h({\bf x}_i)\}]$ under an exponential family representation. Hence, we can write
\begin{eqnarray}
\pi_h({\bf x}_i)=\frac{\exp\{\eta_h({\bf x}_i )\}}{1+\exp\{\eta_h({\bf x}_i )\}} \prod_{l=1}^{h-1}\left[\frac{1}{1+\exp\{\eta_l({\bf x}_i )\}}\right],  \qquad h \in \mathbb{N}, \label{logit_stick}
\end{eqnarray}
allowing each $\eta_h({\bf x}_i )$ to be explicitly interpreted as the log-odds of the probability of being allocated to component $h$, conditionally on the event of surviving to the first $1, \ldots, h-1$ components. This result might be helpful in driving prior specification for the stick-breaking weights, while allowing recent computational advances in Bayesian logistic regression  \citep{PS13} to be inherited in our density regression problem.

To conclude our Bayesian representation, we require priors for the log-odds $\eta_h({\bf{x}}_i)$, $h \in \mathbb{N}$ in the continuation-ratio logistic regressions. A natural choice, which is consistent with classical generalized linear models \citep[e.g.][]{NW72}, is to define $\eta_h({\bf{x}}_i)$ as a linear combination of selected functions of the covariates $\boldsymbol{\psi}({\bf{x}}_i)=\{\psi_1({\bf{x}}_i), \ldots, \psi_R({\bf{x}}_i)\}^{\intercal}$ and consider Gaussian priors for the coefficients, thus obtaining
\begin{eqnarray}
\eta_h({\bf{x}}_i)=\boldsymbol{\psi}({\bf{x}}_i)^\intercal\boldsymbol{\alpha}_h, \quad \mbox{with }\boldsymbol{\alpha}_h \sim \mbox{N}_{R}(\boldsymbol{\mu}_{\alpha}, \boldsymbol{\Sigma}_{\alpha}), \qquad  \qquad h \in \mathbb{N}.
\label{pred}
\end{eqnarray}
Although the linearity assumption in \eqref{pred} may seem restrictive, note that flexible formulations for $\eta_h({\bf{x}}_i)$, including regression via splines and Gaussian processes, induce linear relations  in the coefficients. Moreover, as we will outline in Section~\ref{Section3}, the linearity assumption simplifies computations, while inducing a logistic-normal prior for each $\nu_h({\bf{x}}_i)$. Although such a prior can closely approximate Dirichlet distributions  \citep{AS80}, the logit stick-breaking does not induce beta distributed stick-breaking weights, and therefore it cannot be included in the class discussed by  \citet{IJ01}. However, one can easily adapt the theoretical results in \citet{RD11} to our logit link. For example, the infinite summation of the mixing weights is such that $\sum_{h=1}^\infty \pi_h({\bf x}_i) = 1$ almost surely for any ${\bf x} \in \mathcal{X}$; see the Appendix for details. Moreover, the  \textsc{lsbp} is highly similar in its probabilistic nature and properties to other popular predictor-dependent stick-breaking constructions. In particular,  \textsc{psbp} can be approximated by  \textsc{lsbp}, and viceversa, up to a simple transformation of the prior for each $\boldsymbol{\alpha}_h$. This is a natural consequence of the well known relationship between the probit and the logit function \citep{AM81}, since the mapping $\{1+\exp(-\boldsymbol{\psi}({\bf{x}})^\intercal{\boldsymbol{\alpha}}_h)\}^{-1}$ can be roughly approximated by $\Phi\{\boldsymbol{\psi}({\bf{x}})^\intercal{\boldsymbol{\alpha}}_h\sqrt{\pi/8} \}$. This is summarized in Remark \ref{prop3}.
\begin{remark} 
The logit stick-breaking prior in  \eqref{logit_stick}--\eqref{pred}, can be approximated by a probit stick-breaking process $\nu_h({\bf{x}})\approx \Phi\{\boldsymbol{\psi}({\bf{x}})^\intercal\bar{\boldsymbol{\alpha}}_h\}$, with $\bar{\boldsymbol{\alpha}}_h={\boldsymbol{\alpha}}_h\sqrt{\pi/8} \sim \mbox{\normalfont N}_{R}\{\sqrt{\pi/8}\boldsymbol{\mu}_{\alpha}, (\pi/8)\boldsymbol{\Sigma}_{\alpha}\}$, for every ${\bf{x}} \in \mathcal{X}$ and $h\in \mathbb{N}$.
 \label{prop3}
\end{remark} 
\noindent Hence, a researcher considering a \textsc{psbp} could perform approximate inference leveraging our algorithms, after a suitable rescaling of the prior for each $\boldsymbol{\alpha}_h$. Moreover, this link suggests that the $O(\log n)$ growth of the number of clusters found in empirical studies on the  \textsc{psbp}, should hold also for  \textsc{lsbp}.

\section{Bayesian computational methods}
\label{Section3}

Although the  \textsc{lsbp} and the associated computational procedures apply to a wider set of dependent mixture models and kernels, we focus, for the sake of clarity, on the general class of predictor-dependent infinite mixtures of Gaussians
\begin{eqnarray}
f_{{\bf x}}(y)= \int\sqrt{\tau} \phi[\sqrt{\tau}\left\{y -\boldsymbol{\lambda}({\bf x})^\intercal \boldsymbol{\beta}\right\}]  p_{\bf{x}}( \mbox{d} \boldsymbol{\beta}, \mbox{d}\tau)
=\sum_{h=1}^{\infty} \pi_h({\bf x}) \sqrt{\tau_h} \phi[\sqrt{\tau_h}\left\{y -\boldsymbol{\lambda}({\bf x})^\intercal \boldsymbol{\beta}_h\right\}] ,
\label{mix_dep_gaus}
\end{eqnarray}
where $\tau_h=\sigma^{-2}_h$ is the precision parameter, whereas $\boldsymbol{\beta}_h=(\beta_{1h},\ldots,\beta_{Mh})^\intercal$ denotes a vector of coefficients linearly related to selected functions of the observed predictors $\boldsymbol{\lambda}({\bf x})=\{\lambda_1({\bf x}), \ldots, \lambda_M({\bf x})\}^{\intercal}$. Formulation \eqref{mix_dep_gaus} provides a flexible construction \citep{BAR2012,PA013}, and is arguably the most widely used in Bayesian density regression.  As mentioned in Section \ref{sec1}, we provide here a detailed derivation of three computational methods for Bayesian density regression under model \eqref{mix_dep_gaus}, with logit stick-breaking prior \eqref{logit_stick}--\eqref{pred} for the mixing weights. In particular, we consider a Gibbs sampler converging to the exact posterior, an expectation-maximization (\textsc{em}) algorithm for point estimation, and a mean-field variational Bayes  (\textsc{vb}) approximation for scalable posterior inference. The algorithms associated with these methods are available at \url{https://github.com/tommasorigon/LSBP}, along with the code to reproduce the application in Section \ref{Section4}. 

In the classical predictor-independent mixture of Gaussians framework, these computational methods are closely related, and relevant connections can be drawn also with $k$-means and Bayesian $k$-means algorithms \citep{BI06,Kurihara2009}. A summary of these relations is depicted in Figure 1 of \citet{Kurihara2009}. Broadly speaking, these strategies differ in how they handle unknown parameters and the involved latent quantities, either through maximization or by taking expectations. These connections are paralleled in the \textsc{lsbp} model, although our focus is mainly on Gibbs sampling, \textsc{em} and \textsc{vb}.

Before providing a detailed derivation of these different algorithms, we first study a truncated version of the random probability measure $p_{\bf{x} }$, which will be employed as an approximation of the infinite process. Indeed, although Gibbs samplers for infinite representations are available  \citep{KGW11}, developing \textsc{em} and \textsc{vb} algorithms is not straightforward. In line with  \citet{RD11} and \citet{RDC11}, we develop detailed routines based on a finite representation. In particular, we model the first $H-1$ weights $\nu_1({\bf{x}}), \ldots, \nu_{H-1}({\bf{x}})$ and let $\nu_H({\bf{x}})=1$ for any $ {\bf{x}} \in \mathcal{X}$, so that $\sum_{h=1}^{H} \pi_h({\bf{x}}) = 1$. Based on Theorem~\ref{teo4} below, this choice provides an accurate approximation of the infinite representation for sufficiently large  truncations $H$. 
\begin{theorem}
\label{teo4}
For a sample ${\bf{y}}= (y_1,\dots,y_n)^\intercal$ with covariates ${\bf{X}} = ({\bf{x}}_1, \ldots, {\bf{x}}_n)^\intercal$, let
\begin{eqnarray*}
f_{\bf{X}}^{(H)}({\bf{y}}) = \mbox{\normalfont E}\left( \prod_{i=1}^n \sum_{h=1}^H \pi_h({\bf x}_i) \sqrt{\tau_h} \phi[\sqrt{\tau_h}\left\{y_i -\boldsymbol{\lambda}({\bf x}_i)^\intercal \boldsymbol{\beta}_h\right\}]\right),
\end{eqnarray*}
be the marginal joint density arising from a truncated \textsc{lsbp} prior with $H$ components, and define with $f_{\bf{X}}^{(\infty)}({\bf{y}})$ the same quantity in the infinite case. Moreover, let $\mu_{\nu}({\bf x}) = \mbox{\normalfont E}\{\nu_h({\bf x})\}$ be the expected value of a generic stick-breaking weight. Then $||f_{\bf{X}}^{(H)}({\bf{y}})  - f_{\bf{X}}^{(\infty)}({\bf{y}})||_1  \leq 4\sum_{i=1}^n\{1-\mu_{\nu}({\bf x}_i) \}^{H-1},$ where $|| \cdot ||_1$ denotes the $L^1$--norm. Note that in the above formula the expectation is taken with respect to the \textsc{lsbp} prior law. 
\end{theorem}

According to Theorem  \ref{teo4}, for fixed sample size $n$ and covariates ${\bf{X}}$, the $L^1$ distance between $f_{\bf{X}}^{(H)}({\bf{y}})$ and $f_{\bf{X}}^{(\infty)}({\bf{y}})$ vanishes as $H \rightarrow \infty$, implying that the marginal density $f_{\bf{X}}^{(H)}({\bf{y}})$ converges to $f_{\bf{X}}^{(\infty)}({\bf{y}})$. This rate of decay is exponential in $H$, and therefore the number of components does not have to be very large in practice to accurately approximate the infinite representation, thus motivating computational methods based on truncated versions.

\subsection{MCMC via Gibbs sampling}\label{gibbs}

In deriving a Gibbs sampler for model \eqref{mix_dep_gaus} we focus on a dependent mixture of Gaussians with fixed $H$, and exploit the hierarchical representation~\eqref{marg} along with the continuation-ratio characterization of the logit stick-breaking prior, given in Section \ref{Section2}. Under these constructions, the joint law for the augmented model~\eqref{marg} and its parameters becomes
\begin{eqnarray}
f(\boldsymbol{\alpha}) f(\boldsymbol{\beta}) f(\boldsymbol{\tau})\prod_{i=1}^n\left[\prod_{h=1}^{H}  (\sqrt{\tau_h} \phi[\sqrt{\tau_h}\{y_i -\boldsymbol{\lambda}({\bf x}_i)^\intercal  \boldsymbol{\beta}_h\} ])^{\mathbbm{1}(G_i=h)} \prod_{h=1}^{H-1} \nu_h( {\bf x}_i)^{\mathbbm{1}(G_i = h)}\{ 1-\nu_h( {\bf x}_i)\}^{\mathbbm{1}(G_i > h)} \right],
\label{joint_Gibbs}
\end{eqnarray}
with $\nu_h({\bf x}_i)=\exp\{\psi({\bf x}_i)^{\intercal} \boldsymbol{\alpha}_h \}/[1+\exp\{\psi({\bf x}_i)^{\intercal} \boldsymbol{\alpha}_h \}]$, whereas $f(\boldsymbol{\alpha})f(\boldsymbol{\beta})f(\boldsymbol{\tau}) = \prod_{h=1}^{H-1}f(\boldsymbol{\alpha}_h) \prod_{h=1}^Hf(\boldsymbol{\beta}_h)f({\tau}_h)$ denote the prior laws of the parameters comprising $\boldsymbol{\alpha}$, $\boldsymbol{\beta}$ and $\boldsymbol{\tau}$. As is clear from \eqref{joint_Gibbs}, given  ${\bf{G}}=(G_1, \ldots, G_n)$, sampling of $\boldsymbol{\beta}_h$ and $ \tau_h$, for $h=1, \ldots, H$, requires standard methods for Gaussian linear regression within each mixture component, as long as conditionally conjugate priors $\boldsymbol{\beta}_h \sim \mbox{N}_M(\boldsymbol{\mu}_{\beta},\boldsymbol{\Sigma}_{\beta})$ and $\tau_h \sim \mbox{Ga}(a_{\tau},b_{\tau})$, or normal-gammas for the pair $(\boldsymbol{\beta}_h,\tau_h)$, are employed. Here we focus on the first choice to keep notation  more compact.  

The updating of the $\boldsymbol{\alpha}_h$ parameters, for $h=1, \ldots, H-1$, relies instead on a set of separate Bayesian logistic regressions with responses $z_{ih} = 1$ when $G_i = h$ and $z_{ih} = 0$ if $G_i > h$, for those units $i$ having $G_i > h - 1$, thus allowing parallel sampling from the full-conditional of each $\boldsymbol{\alpha}_h$. Adapting results from the recent P\'olya-gamma data augmentation scheme \citep{PS13} to our statistical model, these updatings can be easily accomplished by noticing that $\nu_h( {\bf x}_i)^{z_{ih}}\{ 1-\nu_h( {\bf x}_i)\}^{1 - z_{ih}} = \int f_{{\bf{x}}_i}(z_{ih})f_{{\bf{x}}_i}(\omega_{ih})\mbox{d}\omega_{ih},$ with laws $f_{{\bf{x}}_i}(z_{ih})$ and $f_{{\bf{x}}_i}(\omega_{ih})$  defined as
\begin{eqnarray}
f_{{\bf{x}}_i}(z_{ih})= \frac{0.5 \exp\{(z_{ih}-0.5)\boldsymbol{\psi}({\bf{x}}_i)^\intercal{\boldsymbol{\alpha}}_h \}}{\mbox{cosh}\{0.5\boldsymbol{\psi}({\bf{x}}_i)^\intercal{\boldsymbol{\alpha}}_h \}},  \quad f_{{\bf{x}}_i}(\omega_{ih})=\frac{\exp[-0.5\{\boldsymbol{\psi}({\bf{x}}_i)^\intercal{\boldsymbol{\alpha}}_h\}^2\omega_{ih} ]f(\omega_{ih})}{[\mbox{cosh}\{0.5\boldsymbol{\psi}({\bf{x}}_i)^\intercal{\boldsymbol{\alpha}}_h \}]^{-1}},
\label{polya}
\end{eqnarray}
for every $i : G_i > h - 1$ and $h=1, \ldots, H-1$. In \eqref{polya}, $f_{{\bf{x}}_i}(\omega_{ih})$ and  $f(\omega_{ih})$ are the density functions of the P\'olya-gamma random variables $\mbox{PG}\{1,\boldsymbol{\psi}({\bf{x}}_i)^\intercal{\boldsymbol{\alpha}}_h\}$, and $\mbox{PG}(1,0)$, respectively. Hence, based on \eqref{polya}, the contribution to the augmented likelihood for each pair $(z_{ih}, \omega_{ih})$ is proportional to a Gaussian kernel for transformed data $(z_{ih}-0.5)/\omega_{ih}$, provided that $f_{{\bf{x}}_i}(z_{ih})f_{{\bf{x}}_i}(\omega_{ih})$ $\propto  \exp[(z_{ih}-0.5)\boldsymbol{\psi}({\bf{x}}_i)^\intercal{\boldsymbol{\alpha}}_h -0.5\{\boldsymbol{\psi}({\bf{x}}_i)^\intercal{\boldsymbol{\alpha}}_h\}^2\omega_{ih}]$. This allows conditionally conjugate updating steps for each $\boldsymbol{\alpha}_h$ under a classical Bayesian linear regression framework. Refer to \citet{CH13,Wang2018b,Wang2018} for further theoretical properties of the P\'olya-gamma scheme. Finally,  note that in~\eqref{polya}, the latent indicators $z_{ih}$ and the P\'olya-gamma random variables $\omega_{ih}$ are conditionally independent given the coefficients $\boldsymbol{\alpha}_h$ for $i : G_i > h-1$. This is in contrast with the data augmentation underlying the \textsc{psbp}, which would lead to more complex calculations, especially in the \textsc{em} and \textsc{vb} algorithms discussed in Sections \ref{em} and \ref{vb}.

\begin{algorithm*}[t!]
\begingroup
    \fontsize{9pt}{10pt}\selectfont
 \caption{Steps of the Gibbs sampler for predictor-dependent finite mixtures of Gaussians} 
 \label{algo1}
     \Begin{\vspace{3pt}
 {\bf [1]}  Assign each unit $i=1, \ldots, n$ to a mixture component $h=1, \ldots, H$\;
 \For(){$i$ \mbox{from} $1$ to $n$}
 {
    Sample  $G_i \in \{1, \ldots, H\}$ from the categorical variable with probabilities $$\mbox{pr}(G_i = h \mid -) = \frac{\left [\nu_h({\bf{x}}_i)\prod_{l=1}^{h-1}\{1 - \nu_l({\bf{x}}_i)\} \right ]\sqrt{\tau_h} \phi[\sqrt{\tau_h}\{y_i -\boldsymbol{\lambda}({\bf{x}}_i)^\intercal  \boldsymbol{\beta}_h\}]}{\sum_{q=1}^H\left [\nu_q({\bf{x}}_i)\prod_{l=1}^{q-1}\{1 - \nu_l({\bf{x}}_i)\} \right ]\sqrt{\tau_q} \phi[\sqrt{\tau_q}\{y_i -\boldsymbol{\lambda}({\bf{x}}_i)^\intercal  \boldsymbol{\beta}_q\}]}, $$
for every $h=1, \ldots, H$.      }
\vspace{3pt}
 {\bf [2]}  Update the parameters ${\boldsymbol{\alpha}}_h$ for $h=1, \ldots, H-1$ exploiting the continuation-ratio representation and the results from the P\'olya-gamma data augmentation in \eqref{polya}\;
 \For(){$h$ \mbox{from} $1$ to $H-1$}
 { \For(){every $i$ \mbox{such that} $G_i>h-1$}
 {Sample the P\'olya-gamma data $\omega_{ih}$ from $(\omega_{ih} \mid -) \sim \mbox{PG}\{1,\boldsymbol{\psi}({\bf{x}}_i)^\intercal{\boldsymbol{\alpha}}_h\}.$} 
 Given  the P\'olya-gamma data, update ${\boldsymbol{\alpha}}_h$ from the full conditional $({\boldsymbol{\alpha}}_h \mid -) \sim \mbox{N}_R(\boldsymbol{\mu}_{\boldsymbol{\alpha}_h},\boldsymbol{\Sigma}_{\boldsymbol{\alpha}_h}),$ having $\boldsymbol{\mu}_{\boldsymbol{\alpha}_h}=\boldsymbol{\Sigma}_{\boldsymbol{\alpha}_h}\{\boldsymbol{\Psi}_h({\bf {x}})^\intercal \boldsymbol{\kappa}_h+ \boldsymbol{\Sigma}_{\boldsymbol{\alpha}}^{-1}\boldsymbol{\mu}_{\boldsymbol{\alpha}}\}$,  $\boldsymbol{\Sigma}_{\boldsymbol{\alpha}_h}=\{\boldsymbol{\Psi}_h({\bf {x}})^\intercal\mbox{diag}(\omega_{1h}, \ldots, \omega_{\bar{n}_h h})\boldsymbol{\Psi}_h({\bf {x}})+\boldsymbol{\Sigma}_{\boldsymbol{\alpha}}^{-1}\}^{-1}$,  where $\boldsymbol{\kappa}_h=(z_{1h}-0.5, \ldots, z_{\bar{n}_h h}-0.5)^\intercal$, with $z_{ih}=1$ if $G_i=h$ and $z_{ih}=0$ if $G_i>h$. }
\vspace{3pt}
{\bf [3]} Update the kernel parameters ${\boldsymbol{\beta}}_h$, $h=1, \ldots, H$, in \eqref{mix_dep_gaus}, leveraging standard Bayesian linear regression\;
 \For(){$h$ \mbox{from} $1$ to $H$}
 {
    Sample the coefficients comprising  ${\boldsymbol{\beta}}_h$ from the full conditional   $({\boldsymbol{\beta}}_h \mid -) \sim \mbox{N}_M(\boldsymbol{\mu}_{\boldsymbol{\beta}_h},\boldsymbol{\Sigma}_{\boldsymbol{\beta}_h})$, with $\boldsymbol{\mu}_{\boldsymbol{\beta}_h}=\boldsymbol{\Sigma}_{\boldsymbol{\beta}_h}\{\tau_h \boldsymbol{\Lambda}_h({\bf {x}})^\intercal {\bf{y}}_h+ \boldsymbol{\Sigma}_{\boldsymbol{\beta}}^{-1}\boldsymbol{\mu}_{\boldsymbol{\beta}}\}$,  $\boldsymbol{\Sigma}_{\boldsymbol{\beta}_h}=\{\tau_h \boldsymbol{\Lambda}_h({\bf {x}})^\intercal \boldsymbol{\Lambda}_h({\bf {x}})+\boldsymbol{\Sigma}_{\boldsymbol{\beta}}^{-1}\}^{-1}$, and ${\bf{y}}_h$ the $n_h \times 1$ vector containing the responses for all the units with $G_i=h$.}
\vspace{3pt}
{\bf [4]} Update the precision parameters $\tau_h$, $h=1, \ldots, H$ of each kernel in \eqref{mix_dep_gaus}\;
 \For(){$h$ \mbox{from} $1$ to $H$}
 {
Sample $\tau_h$ from $(\tau_h \mid -) \sim \mbox{Ga}[a_\tau +  0.5\sum_{i=1}^n\mathbbm{1}(G_i=h), b_\tau + 0.5\sum_{i : G_i=h}\{y_i - \boldsymbol{\lambda}({\bf x}_i)^\intercal  \boldsymbol{\beta}_h\}^2]$.}
    }
    \endgroup
\end{algorithm*}

The detailed steps of the Gibbs sampler for the truncated representation of model   \eqref{mix_dep_gaus} are outlined in Algorithm \ref{algo1}. In this routine, $\boldsymbol{\Lambda}_h({\bf {x}})$ and $\boldsymbol{\Psi}_h({\bf {x}})$ denote the $n_h \times M$ and the $\bar{n}_h \times R$ predictor matrices in  \eqref{mix_dep_gaus} and  \eqref{pred} having row entries $\boldsymbol{\lambda}({\bf x}_i)^\intercal$ and $\boldsymbol{\psi}({\bf x}_i)^\intercal$, for only those statistical units $i$ such that $G_i =h$ and $G_i > h-1$, respectively. We shall also emphasize that step {\bf [1]} can be run in parallel across units $i=1, \ldots, n$, whereas parallel computing for the different mixture components can be easily implemented in steps  {\bf [2]}, {\bf [3]} and {\bf [4]}.

\subsection{EM algorithm} \label{em}
In high-dimensional studies, the Gibbs sampler described in Section \ref{gibbs} could face computational bottlenecks. If a point estimate of model \eqref{mix_dep_gaus} is the main quantity of interest, for example for prediction purposes, one possibility is to rely on a more efficient procedure specifically designed for this goal, such as the  \textsc{em} \citep{DL77}. The implementation of a simple \textsc{em} for a finite representation of model \eqref{mix_dep_gaus} under the \textsc{lsbp} prior benefits from the P\'olya-gamma data augmentation, which has analytical expectation and allows direct maximization within a Gaussian linear regression framework. Note that, although the \textsc{em} algorithm is commonly implemented for maximum likelihood estimation, it can be easily modified to estimate posterior modes  \citep[e.g.][]{DL77}. 

\begin{algorithm*}[t!]
\begingroup
    \fontsize{9pt}{11pt}\selectfont
 \caption{Steps of the \textsc{em} algorithm  for predictor-dependent finite mixtures of Gaussians} 
 \label{algo2}
 
     \Begin{\vspace{3pt}
      Let $( \boldsymbol{\alpha}^{(t)}$, $\boldsymbol{\beta}^{(t)}$, $\boldsymbol{\tau}^{(t)})$ denote the values of the parameters at iteration $t$.
      
      \vspace{5pt}
 {\bf [1] Expectation:}   Exploiting results in \eqref{EM}, the expectation of \eqref{eq::completeloglik} with respect to the augmented data $(\boldsymbol{\zeta}_i, \bar{\boldsymbol{\omega}}_i)$, for each $i=1, \ldots, n$, can be obtained by plugging in $\hat{\boldsymbol{\zeta}}_i=\mbox{E}(\boldsymbol{\zeta}_i\mid {{y}}_i, {\bf{x}}_i, \boldsymbol{\beta}^{(t)}, \boldsymbol{\tau}^{(t)})$ and $\hat{\bar{\boldsymbol{\omega}}}_i=\mbox{E}(\bar{{\boldsymbol{\omega}}}_i\mid  {\bf{x}}_i,\hat{\boldsymbol{\zeta}}_i, \boldsymbol{\alpha}^{(t)})$ in \eqref{EM}. Therefore:
 
 \For(){$i$ \mbox{from} $1$ to $n$}
 {\For(){$h$ \mbox{from} $1$ to $H$}
 {Compute $\hat{{{\zeta}}}_{ih}$ by applying the following expression $$\hat{{{\zeta}}}_{ih} = \frac{\left [\nu_h^{(t)}({\bf{x}}_i)\prod_{l=1}^{h-1}\{1 - \nu_l^{(t)}({\bf{x}}_i)\} \right ]\sqrt{\tau^{(t)}_h} \phi[\sqrt{\tau^{(t)}_h}\{y_i -\boldsymbol{\lambda}({\bf{x}}_i)^\intercal  \boldsymbol{\beta}^{(t)}_h\}]}{\sum_{q=1}^H\left [\nu_q^{(t)}({\bf{x}}_i)\prod_{l=1}^{q-1}\{1 - \nu_l^{(t)}({\bf{x}}_i)\} \right ]\sqrt{\tau^{(t)}_q} \phi[\sqrt{\tau^{(t)}_q}\{y_i -\boldsymbol{\lambda}({\bf{x}}_i)^\intercal  \boldsymbol{\beta}^{(t)}_q\}]},$$ and calculate $\hat{\bar{{\omega}}}_{ih}$ via $\hat{\bar{{\omega}}}_{ih} =\{2\boldsymbol{\psi}({\bf{x}}_i)^\intercal\boldsymbol{\alpha}_h^{(t)}\}^{-1}\tanh{\{0.5\boldsymbol{\psi}({\bf{x}}_i)^\intercal\boldsymbol{\alpha}_h^{(t)}\}}\sum_{l=h}^H\hat{\zeta}_{il}$, \citep{PS13}.} }
\vspace{3pt}
 {\bf [2] Maximization:}  To maximize the expected complete log-posterior $\mbox{log}f_{\bf{x}} (\boldsymbol{\alpha}, \boldsymbol{\beta}, \boldsymbol{\tau}\mid {\bf{y}},\hat{\boldsymbol{\zeta}},\hat{\bar{\boldsymbol{\omega}}} )$, note that according to   \eqref{eq::completeloglik}--\eqref{EM}, modes $\boldsymbol{\alpha}^{(t+1)}$ and $(\boldsymbol{\beta}^{(t+1)},\boldsymbol{\tau}^{(t+1)} )$ can be obtained separately as follow:
 
 \For(){$h$ \mbox{from} $1$ to $H-1$}
 {To compute $\boldsymbol{\alpha}_h^{(t+1)}$, note that since $\boldsymbol{\alpha}_h$ has Gaussian prior, and provided that the second term in \eqref{EM} is based on Gaussian kernels, the estimated  $\boldsymbol{\alpha}_h$ at step $t+1$ coincides with the mean of a full conditional Gaussian, similar to the one in step {\bf{[2]}} of Algorithm \ref{algo1}.
 $$\boldsymbol{\alpha}_h^{(t+1)}=\{\boldsymbol{\Psi}({\bf {x}})^\intercal \mbox{diag}(\hat{\bar{\omega}}_{1h}, \ldots, \hat{\bar{\omega}}_{nh})\boldsymbol{\Psi}({\bf {x}})+\boldsymbol{\Sigma}_{\boldsymbol{\alpha}}^{-1}\}^{-1}\{\boldsymbol{\Psi}({\bf {x}})^\intercal (\hat{\bar{\kappa}}_{1h}, \ldots, \hat{\bar{\kappa}}_{nh})^\intercal+ \boldsymbol{\Sigma}_{\boldsymbol{\alpha}}^{-1}\boldsymbol{\mu}_{\boldsymbol{\alpha}}\}, $$
 where each $\hat{\bar{\kappa}}_{ih} = \hat{\zeta}_{ih} - 0.5\sum_{l=h}^H\hat{\zeta}_{il}$ and $\boldsymbol{\Psi}({\bf {x}})$ is the design matrix of the logistic regression based on all units.} 
  \For(){$h$ \mbox{from} $1$ to $H$}
 {A similar approach can be considered to compute  $\boldsymbol{\beta}_h^{(t+1)}$ and $\tau_h^{(t+1)}$ under the Gaussian and gamma priors for these parameters and the Gaussian kernel characterizing the first term in \eqref{EM}. Hence,  adapting steps {\bf{[3]}} and {\bf{[4]}} in Algorithm \ref{algo1} to the \textsc{em} setting, provides:
\begin{eqnarray*}
\boldsymbol{\beta}_h^{(t+1)}&{=}&\{\tau^{(t)}_{h}\boldsymbol{\Lambda}({\bf {x}})^\intercal  \mbox{diag}(\hat{\zeta}_{1h}, \ldots, \hat{\zeta}_{nh})\boldsymbol{\Lambda}({\bf {x}})+\boldsymbol{\Sigma}_{\boldsymbol{\beta}}^{-1}\}^{-1}\{\tau^{(t)}_{h}\boldsymbol{\Lambda}({\bf {x}})^\intercal\mbox{diag}(\hat{\zeta}_{1h}, \ldots, \hat{\zeta}_{nh}){\bf{y}}+ \boldsymbol{\Sigma}_{\boldsymbol{\beta}}^{-1}\boldsymbol{\mu}_{\boldsymbol{\beta}}\},  \\
\tau_h^{(t+1)}&{=}&\mbox{max}\{0, [a_{\tau}+0.5\sum_{i=1}^n \hat{\zeta}_{ih}-1][b_{\tau}+0.5 \sum_{i=1}^n \hat{\zeta}_{ih}\{y_i-\boldsymbol{\lambda}({\bf{x}}_i)^\intercal \boldsymbol{\beta}_h^{(t)}\}^2]^{-1}\},
\end{eqnarray*}
\vspace{-10pt}

where $\boldsymbol{\Lambda}({\bf {x}})$ is the design matrix of the Gaussian regression within each kernel based on all units.} }
\endgroup
\end{algorithm*}

The proposed \textsc{em} in Algorithm \ref{algo2} alternates between a maximization step for the parameters $(\boldsymbol{\alpha}, \boldsymbol{\beta},  \boldsymbol{\tau})$ and an expectation step for the augmented data  $(\boldsymbol{\zeta}_i,\bar{\boldsymbol{\omega}}_i)$, $i=1, \ldots, n$, with $\boldsymbol{\zeta}_{i}=\{\zeta_{i1}=\mathbbm{1}(G_i=1), \ldots, \zeta_{iH}=\mathbbm{1}(G_i=H) \}^\intercal$ the vector of binary indicators denoting the membership to a mixture component, and  $\bar{\boldsymbol{\omega}}_i=(\bar{\omega}_{i1}, \ldots, \bar{\omega}_{iH-1})^{\intercal}$ the corresponding P\'olya-gamma augmented data. Although this data augmentation parallels the one described for the Gibbs sampler, we adopt a slightly different notation for the P\'olya-gamma random variables $\bar{\omega}_{ih}$, to emphasize that we are considering $n$ units, and not only those for which the cluster indicators $G_i > h - 1$. Indeed, in line with the \textsc{em} rationale, we do not condition on the membership indicators and on the P\'olya-gamma latent random variables, but we rather take expectations with respect to their conditional distributions. For the same reason, in this case we work directly with the component indicator variables $\boldsymbol{\zeta}_{i}$ instead of the binary vectors ${\bf{z}}_i=(z_{i1}, \ldots, z_{iH-1})^{\intercal}$ in \eqref{bin_seq}. 

Based on the data augmentations outlined in \eqref{marg} and \eqref{polya}, the complete log-posterior $\mbox{log} f_{{\bf{x}}}(\boldsymbol{\alpha}, \boldsymbol{\beta}, \boldsymbol{\tau}\mid {\bf{y}},\boldsymbol{\zeta},\bar{\boldsymbol{\omega}})$ underlying the proposed \textsc{em} routine, can be written as
\begin{eqnarray}
\label{eq::completeloglik}
 \sum_{i=1}^n\ell_{{\bf{x}}_i }(\boldsymbol{\alpha}, \boldsymbol{\beta},  \boldsymbol{\tau}; {y}_i,\boldsymbol{\zeta}_i,\bar{\boldsymbol{\omega}}_i) +\sum_{h=1}^{H-1}  \log{f(\boldsymbol{\alpha}_h)  +\sum_{h=1}^{H}\log f(\boldsymbol{\beta}_h)+ \sum_{h=1}^{H}\log f({\tau}_h)}+ \mbox{const},
\end{eqnarray}
where $\ell_{{\bf{x}}_i }(\boldsymbol{\alpha}, \boldsymbol{\beta},  \boldsymbol{\tau}; {y}_i,\boldsymbol{\zeta}_i,\bar{\boldsymbol{\omega}}_i)$ is the contribution of unit $i$ to the complete log-likelihood. Working on the complete log-likelihood has relevant benefits. Indeed, exploiting  equations \eqref{marg} and \eqref{bin_seq}, and the results in  \citet{PS13} summarized in \eqref{polya}, the term $\ell_{{\bf{x}}_i }(\boldsymbol{\alpha}, \boldsymbol{\beta},  \boldsymbol{\tau}; {y}_i,\boldsymbol{\zeta}_i,\bar{\boldsymbol{\omega}}_i)=\ell_{{\bf{x}}_i }( \boldsymbol{\beta},  \boldsymbol{\tau}; {y}_i,\boldsymbol{\zeta}_i)+\ell_{{\bf{x}}_i }{(\boldsymbol{\alpha}; \boldsymbol{\zeta}_i,\bar{\boldsymbol{\omega}}_i )}$, can be factorized as
\begin{eqnarray}
\sum_{h=1}^H\zeta_{ih}\left[-\frac{\tau_h\{y_i -\boldsymbol{\lambda}({\bf x}_i)^\intercal  \boldsymbol{\beta}_h\}^2}{2} +\frac{1}{2}\log(\tau_h)\right]+ \sum_{h=1}^{H-1}\left[\bar{\kappa}_{ih}\boldsymbol{\psi}({\bf{x}}_i)^\intercal\boldsymbol{\alpha}_h -\bar{\omega}_{ih}\frac{\{\boldsymbol{\psi}({\bf{x}}_i)^\intercal\boldsymbol{\alpha}_h\}^2}{2}\right]+ \mbox{const},
\label{EM}
\end{eqnarray}
where $\bar{\kappa}_{ih}=\zeta_{ih}-0.5\sum_{l=h}^H\zeta_{il}$. Hence, both terms in equation \eqref{EM} are linear in the augmented data $(\boldsymbol{\zeta}_i, \bar{\boldsymbol{\omega}}_i)$, and represent the sum of Gaussian kernels. This linearity property  simplifies computations in the expectation step for the complete log-posterior in equation \eqref{eq::completeloglik}, whereas the Gaussian structure allows simple maximizations.  Since the joint maximization of the expected complete log-posterior with respect to $(\boldsymbol{\beta},\boldsymbol{\tau})$  is intractable, we rely on a conditional maximization procedure \citep{MR93} in the last step of Algorithm \ref{algo2}, which provides analytical solutions.

\subsection{Mean-field variational Bayes}\label{vb}

Section \ref{em} provides a scalable procedure for estimation of posterior modes in large-scale problems. However, an appealing aspect of the Bayesian approach is in allowing uncertainty quantification via inference on the entire posterior. The Gibbs sampler in Section~\ref{gibbs} represents an appealing procedure which converges to the exact posterior, but faces computational bottlenecks. This motivates scalable variational methods for approximate Bayesian inference \citep{BI06,Blei2017}. Clearly, these computational gains do not come without some drawbacks. For example, variational approximations typically underestimate posterior variability. This issue might be mitigated via a post-processing operation as in \citet{Giordano2015}, at the cost of an additional computational step. 

Due to the P\'olya-gamma data augmentation, our variational strategy is framed within the well-established exponential family setting, for which there exists a closed-form coordinate ascent variational inference algorithm (\textsc{cavi}). Compared to more accurate black-box variational strategies \citep[e.g.][]{Ranganath2014}, the \textsc{cavi} algorithm is appealing because it requires no tuning. Moreover, recent theoretical properties for this class of computational methods \citep{Blei2017} are inherited by our variational algorithm. This seems in contrast with the variational strategy discussed by \citet{RDC11}, which considers a local approximation based on the lower bound of \citet{JJ00}. However, the recent contribution of \citet{Durante2019} allows to draw a sharp connection between the P\'olya-gamma data augmentation and the \citet{JJ00} lower bound. As a consequence, the \textsc{vb} approach we propose relies on the same optimization problem considered by \citet{RDC11}.

\begin{algorithm*}[t!]
\begingroup
    \fontsize{9pt}{11pt}\selectfont
 \caption{Steps of the \textsc{cavi} algorithm  for predictor-dependent finite mixtures of Gaussians} 
 \label{algo3}
\Begin{\vspace{3pt}
Let $q^{(t)}(\cdot)$ denote the generic variational distribution at iteration $t$.

\vspace{0pt}
{\bf [1]} Compute $q_{{\bf{x}}_i}^{*(t)}(z_{ih})$, for each $i=1, \ldots, n$ and $h=1, \ldots, H-1$;

 \For(){$i$ from $1$ to $n$}{
\For(){$h$  from $1$ to $H-1$}{
\vspace{2pt}
It can be easily shown that the optimal solution $q_{{\bf{x}}_i}^{*(t)}(z_{ih})$ for the variational distribution of each $z_{ih}$ coincides with  the probability mass function of a $\mbox{Bern}(\rho_{ih})$, having 
\begin{eqnarray*}
\mbox{logit}(\rho_{ih}) = \boldsymbol{\psi}({\bf{x}}_i)^\intercal\mbox{E}(\boldsymbol{\alpha}_h) + \sum_{l=h}^H \zeta_{il}^{(h)}\left[0.5\cdot\mbox{E}(\log{\tau_l}) -0.5\cdot\mbox{E}(\tau_l)\mbox{E}\{(y_i - \boldsymbol{\lambda}({\bf{x}}_i)^\intercal  \boldsymbol{\beta}_l)^2\}\right], 
\end{eqnarray*}
where the expectations are taken with the respect to the current variational distributions for the other parameters, whereas $\zeta^{(h)}_{il}=\prod_{r=1}^{l-1}(1-\rho_{ir})$ if $l=h$, and $\zeta^{(h)}_{il}=-\rho_{il}\prod_{r=1,r\neq h}^{l-1}(1-\rho_{ir})$ otherwise. Note also that $\rho_{iH}=1$.   }}

 {\bf [2]} Compute $q^{*(t)}_{{\bf{x}}}(\boldsymbol{\alpha}_{h})$, for each $h=1, \ldots, H-1$;

 \For(){$h$ \mbox{from} $1$ to $H-1$}{
  \vspace{0pt}
  
  The optimal solution $q^{*(t)}_{{\bf{x}}}(\boldsymbol{\alpha}_h)$ for the variational distribution of each $\boldsymbol{\alpha}_h$ is the density of the Gaussian random variable $\mbox{N}_R[\{\boldsymbol{\Psi}({\bf{x}})^\intercal{\bf{V}}_h\boldsymbol{\Psi}({\bf{x}}) + \boldsymbol{\Sigma}_{\alpha}^{-1}\}^{-1}\{\boldsymbol{\Psi}({\bf{x}})^\intercal \boldsymbol{\rho}_h + \boldsymbol{\Sigma}_{\alpha}^{-1}\boldsymbol{\mu}_{\alpha}\},\{\boldsymbol{\Psi}({\bf{x}})^\intercal {\bf{V}}_h\boldsymbol{\Psi}({\bf{x}}) + \boldsymbol{\Sigma}_{\alpha}^{-1}\}^{-1}]$ with ${\bf{V}}_h {=} \ \mbox{diag}\{\mbox{E}(\omega_{1h}),\dots,\mbox{E}(\omega_{nh})\}$ and $ \boldsymbol{\rho}_h{=}\ (\rho_{1h} - 0.5,\dots,\rho_{nh} - 0.5)^{\intercal}$.}
  \vspace{3pt}
   {\bf [3]} Compute the variational distribution $q^{*(t)}_{{\bf{x}}_i}(\omega_{ih})$ for each $i=1, \ldots, n$ and $h=1, \ldots, H-1$; 
 
 \For(){$i$ from $1$ to $n$}{
\For(){$h$  from $1$ to $H-1$}{
\vspace{0pt}
Update the optimal solution $q^{*(t)}_{{\bf{x}}_i}(\omega_{ih})$ to obtain the density of a P\'olya-gamma $\text{PG}\left(1, \xi_{ih}\right)$, with $\xi_{ih}^2 = \boldsymbol{\psi}({\bf{x}}_i)^\intercal \mbox{E}(\boldsymbol{\alpha}_h\boldsymbol{\alpha}_h^\intercal )\boldsymbol{\psi}({\bf{x}}_i).$
Recall that $\mbox{E}(\omega_{ih}) = 0.5\xi_{ih}^{-1}\tanh(0.5 \xi_{ih})$. }}

\vspace{3pt}
 {\bf [4]} Compute $q^{*(t)}_{{\bf{x}}}(\boldsymbol{\beta}_{h})$ and  $q^{*(t)}_{{\bf{x}}}({\tau}_{h})$, for each $h=1, \ldots, H$;

  \For(){$h$ \mbox{from} $1$ to $H$}{
   \vspace{0pt}
   
Update variational solutions for $\boldsymbol{\beta}_h$ and $\tau_h$. In particular, $q^{*(t)}_{{\bf{x}}}(\boldsymbol{\beta}_{h})$ and $q^{*(t)}_{{\bf{x}}}({\tau}_{h})$ are easily available as the densities of the Gaussian $\mbox{N}_M[\{\boldsymbol{\Lambda}({\bf {x}})^\intercal {\boldsymbol{\Gamma}}_h\boldsymbol{\Lambda}({\bf {x}})+\boldsymbol{\Sigma}_{\boldsymbol{\beta}}^{-1}\}^{-1}\{\boldsymbol{\Lambda}({\bf {x}})^\intercal{\boldsymbol{\Gamma}}_h{\bf{y}}+ \boldsymbol{\Sigma}_{\boldsymbol{\beta}}^{-1}\boldsymbol{\mu}_{\boldsymbol{\beta}}\},\{\boldsymbol{\Lambda}({\bf {x}})^\intercal {\boldsymbol{\Gamma}}_h\boldsymbol{\Lambda}({\bf {x}})+\boldsymbol{\Sigma}_{\boldsymbol{\beta}}^{-1}\}^{-1}] $ and the gamma $\mbox{Ga}[a_{\tau}+0.5\sum_{i=1}^n \mbox{E}({\zeta}_{ih}), b_{\tau}+0.5 \sum_{i=1}^n \mbox{E}({\zeta}_{ih})\mbox{E}\{y_i-\boldsymbol{\lambda}({\bf{x}}_i)^\intercal \boldsymbol{\beta}_h\}^2]$, respectively, with ${\boldsymbol{\Gamma}}_h=\mbox{E}(\tau_{h}) \mbox{diag}\{\mbox{E}({\zeta}_{1h}), \ldots, \mbox{E}({\zeta}_{nh})\}$ and ${\zeta}_{ih}=z_{ih} \prod_{l=1}^{h-1}(1-z_{il})$, $i=1, \ldots, n$.
}   
}
\endgroup
\end{algorithm*}

Compared to the Gibbs sampler in Section~\ref{gibbs}, here we augment the entire model~\eqref{mix_dep_gaus} with respect to the binary vectors ${ \bf z}_i = (z_{i1},\dots,z_{iH-1})^\intercal$, $i=1,\dots,n$ comprising ${ \bf z}$, rather than using the membership indicators ${\bf G}$. Hence, the joint law $f_{\bf{x}}({\bf{y}}, \boldsymbol{\alpha}, \boldsymbol{\beta}, \boldsymbol{\tau},{\bf{z}}, \boldsymbol{\omega})= f_{\bf{x}}({\bf{y}} \mid  {\bf{z}}, \boldsymbol{\beta}, \boldsymbol{\tau})f_{\bf{x}}({\bf{z}} \mid  \boldsymbol{\alpha})f_{\bf{x}}(\boldsymbol{\omega} \mid \boldsymbol{\alpha})f(\boldsymbol{\alpha})f(\boldsymbol{\beta})f(\boldsymbol{\tau})$ is equal to
\begin{eqnarray}
f(\boldsymbol{\alpha}) f(\boldsymbol{\beta}) f(\boldsymbol{\tau})\prod_{i=1}^n\left[\prod_{h=1}^{H} (\sqrt{\tau_h} \phi[\sqrt{\tau_h}\{y_i -\boldsymbol{\lambda}({\bf x}_i)^\intercal  \boldsymbol{\beta}_h\} ])^{z_{ih}\prod_{l=1}^{h-1}(1 - z_{il})}\prod_{h=1}^{H-1} \frac{f(\omega_{ih})}{2} \frac{\exp\{(z_{ih} - 0.5)\boldsymbol{\psi}({\bf{x}}_i)^\intercal\boldsymbol{\alpha}_h\}}{\exp{\{0.5\omega_{ih}(\boldsymbol{\psi}({\bf{x}}_i)^\intercal\boldsymbol{\alpha}_h)^2\}}}\right],
\label{joint_z}
\end{eqnarray}
where $z_{iH} = 1$. Our goal is to find a variational distribution $q_{{\bf{x}}}(\boldsymbol{\alpha},\boldsymbol{\beta}, \boldsymbol{\tau}, {\bf{z}}, \boldsymbol{\omega})$ that best approximates the joint posterior $f_{{\bf{x}}}( \boldsymbol{\alpha}, \boldsymbol{\beta}, \boldsymbol{\tau}, {\bf z }, \boldsymbol{\omega} \mid {\bf y})$, while maintaining simple computations. This can be obtained by minimizing the Kullback-Leibler divergence $\textsc{kl}\{q_{{\bf{x}}}(\boldsymbol{\alpha},\boldsymbol{\beta}, \boldsymbol{\tau}, {\bf{z}}, \boldsymbol{\omega})\mid\mid f_{{\bf{x}}}(\boldsymbol{\alpha},\boldsymbol{\beta}, \boldsymbol{\tau}, {\bf{z}}, \boldsymbol{\omega} \mid {\bf{y}})\}$ between the variational distribution and the full posterior, or, alternatively, by maximizing the evidence lower bound $\textsc{elbo}\{q_{{\bf{x}}}(\boldsymbol{\alpha}, \boldsymbol{\beta}, \boldsymbol{\tau}, {\bf{z}}, \boldsymbol{\omega} ) \}$ of the log-marginal density $\log f_{{\bf{X}}}^{(H)}({\bf{y}})$, since $\log f_{{\bf{X}}}^{(H)}({\bf{y}})$ can be analytically expressed  as the sum of the \textsc{elbo} and the positive \textsc{kl} divergence. This evidence lower bound  to be maximized can be expressed as
\begin{eqnarray*}
\sum_{\bf{z}} \int q_{ {\bf{x}}}(\boldsymbol{\alpha},\boldsymbol{\beta}, \boldsymbol{\tau}, {\bf{z}}, \boldsymbol{\omega}) \left[\log{\left\{  \frac{ f_{\bf{x}}({\bf{y}} \mid  {\bf{z}}, \boldsymbol{\beta}, \boldsymbol{\tau})f_{\bf{x}}({\bf{z}} \mid  \boldsymbol{\alpha})f_{\bf{x}}(\boldsymbol{\omega} \mid \boldsymbol{\alpha})f(\boldsymbol{\alpha})f(\boldsymbol{\beta})f(\boldsymbol{\tau})}{q_{ {\bf{x}}}(\boldsymbol{\alpha},\boldsymbol{\beta}, \boldsymbol{\tau}, {\bf{z}}, \boldsymbol{\omega})}\right\}} \right] \mbox{d}(\boldsymbol{\alpha},\boldsymbol{\beta}, \boldsymbol{\tau}, \boldsymbol{\omega}).\end{eqnarray*}
Without further restrictions, the Kullback-Leibler divergence is minimized when the variational distribution is equal to the true posterior, which is intractable. To address this issue, a common strategy  is to assume that the variational distribution belongs to a mean-field family \citep[see e.g.][]{Blei2017}. This incorporates a posteriori independence among distinct groups of parameters, implying that the variational distribution  can be expressed as the product of marginal laws. Specifically, we consider the following factorization for the variational distribution
\begin{eqnarray}
\label{mean_field}
q_{\bf{x}}(\boldsymbol{\alpha},\boldsymbol{\beta}, \boldsymbol{\tau}, {\bf{z}}, \boldsymbol{\omega}) = \prod_{h=1}^{H-1}q_{\bf{x}}(\boldsymbol{\alpha}_h)\prod_{h=1}^{H}q_{\bf{x}}(\boldsymbol{\beta}_h)\prod_{h=1}^{H}q_{\bf{x}}({\tau}_h) \prod_{h=1}^{H-1}\prod_{i=1}^n q_{{\bf{x}}_i}(z_{ih})\prod_{h=1}^{H-1}\prod_{i=1}^n q_{{\bf{x}}_i}(\omega_{ih}).
\end{eqnarray}
Note that we are not making specific assumptions about the functional form of the variational distributions. Combining \eqref{joint_z} with \eqref{mean_field}, we obtain a tractable expression for the \textsc{elbo}, which can be easily maximized as in \citet[Ch. 10]{BI06}. In particular, the optimal solutions are provided by the following system of equations
\begin{eqnarray*}
\begin{aligned}
&  \log{q_{{\bf{x}}}^{*}(\boldsymbol{\beta}_{h})}=\mbox{E}_{\boldsymbol{\tau},{\bf{z}}}[\log\{   f_{{\bf{x}}}({\bf{y}} \mid  {\bf{z}}, \boldsymbol{\beta}, \boldsymbol{\tau})f(\boldsymbol{\beta}_h)\}] + c_{\boldsymbol{\beta}_h}, \qquad \ \ \ \ \ h=1, \ldots, H, \\
&  \log{q_{{\bf{x}}}^{*}({\tau}_{h})}=\mbox{E}_{\boldsymbol{\beta},{\bf{z}}}[\log\{   f_{{\bf{x}}}({\bf{y}} \mid  {\bf{z}}, \boldsymbol{\beta}, \boldsymbol{\tau})f({\tau}_h)\}] + c_{{\tau}_h}, \qquad \quad \ \ \  h=1, \ldots, H, \\
& \log{q_{{\bf{x}}}^{*}(\boldsymbol{\alpha}_{h})}=\mbox{E}_{ {\bf{z}},\boldsymbol{\omega}}[\log\{  f_{{\bf{x}}}({\bf{z}},\boldsymbol{\omega} \mid  \boldsymbol{\alpha}) f(\boldsymbol{\alpha}_h)\}] + c_{\boldsymbol{\alpha}_h}, \qquad \quad \ \ \ h=1, \ldots, H-1, \\
& \log{q_{{\bf{x}}_i}^{*}(z_{ih})}=\mbox{E}_{\boldsymbol{\alpha},\boldsymbol{\beta},\boldsymbol{\tau}, {\bf{z}}_{i,-h}}[\log  f_{{\bf{x}}}({\bf{y}}, {\bf{z}} \mid \boldsymbol{\beta}, \boldsymbol{\tau},  \boldsymbol{\alpha})] + c_{z_{ih}}, \quad \ \ \ h=1, \ldots, H-1, \ i=1, \ldots, n, \\
&  \log{q_{{\bf{x}}_i}^{*}(\omega_{ih})}=\mbox{E}_{\boldsymbol{\alpha}}[\log f_{{\bf{x}}}(\omega_{ih} \mid \boldsymbol{\alpha})] + c_{\omega_{ih}},  \qquad \ \ \quad  \qquad  \qquad  h=1, \ldots, H-1, \ i=1, \ldots, n,
    \end{aligned}
\end{eqnarray*}
where ${\bf{z}}_{i,-h}$ denotes the vector of binary indicators ${\bf z}_i$ without considering the $h$th one, whereas $c_{\boldsymbol{\beta}_h}$, $c_{{\tau}_h}$, $c_{\boldsymbol{\alpha}_h}$, $c_{z_{ih}}$ and $c_{\omega_{ih}}$, are additive constants with respect to the argument in the corresponding variational distribution. Each expectation in the above equations is evaluated with respect to the variational distribution of the other parameters, and therefore we need to rely on iterative methods to find the optimal solution. We consider the coordinate ascent variational inference (\textsc{cavi}) iterative procedure---described  in Algorithm \ref{algo3}---which maximizes the variational distribution of each parameter based on the current estimate for the remaining ones \citep[e.g.][Ch. 10]{BI06}. This procedure generates a monotone sequence for the $\textsc{elbo}\{q_{ {\bf{x}}}(\boldsymbol{\alpha},\boldsymbol{\beta}, \boldsymbol{\tau}, {\bf{z}}, \boldsymbol{\omega}) \}$, which ensures convergence to a local joint maximum. Refer to \citet{Blei2017} for practical guidelines to address issues of local maxima via multiple runs. Finally, note that as shown in Algorithm \ref{algo3}, the normalizing constants in the above equations have not to be computed numerically, since kernels of well known distributions can be recognized.

\section{Epidemiology application}
\label{Section4}
We compare the performance of the three computational methods developed in Section \ref{Section3}, in a toxicology study. Consistent with recent interests in Bayesian density regression \citep[e.g.][]{DP08,HWA2014,Canale2018}, we focus on a dataset aimed at studying the relationship between the \texttt{DDE} concentration in maternal serum, and the gestational days at delivery \citep{LO01}. 

The \texttt{DDE} is a  metabolite of \textsc{ddt}, which is still used against malaria-transmitting mosquitoes in certain developing countries---according to the Malaria Report 2015 from the World Health Organization---thus raising concerns about its adverse effects on premature delivery. Popular studies in reproductive epidemiology address this goal by dichotomizing the gestational age at delivery (\texttt{GAD}) with a clinical threshold, so that births occurred before the $37$th week are considered preterm. Although this approach allows for a simpler modeling strategy, it leads to a clear loss of information. In particular, a greater risk of mortality and morbidity is associated with preterm birth, which increases rapidly as the \texttt{GAD} decreases. This has motivated an increasing interest in modeling how the entire distribution of \texttt{GAD} changes with \texttt{DDE} exposure \citep[e.g.][]{DP08,HWA2014,Canale2018}.

Data are composed by $n=2312$ measurements $(x_i, y_i)$, $i=1,\dots,n$, where $x_i$ denotes the \texttt{DDE} concentration, and $y_i$ is the gestational age at delivery for woman $i$. Our goal is to reproduce the analyses in \citet{DP08} on this dataset, and compare the inference and computational performance of the \textsc{mcmc} via Gibbs sampling, the \textsc{em} algorithm, and the  \textsc{vb} routine proposed in Section \ref{Section3}. Note that, consistent with the main novelty of this contribution, we do not attempt to improve the flexibility and the efficiency of the available statistical models for Bayesian density regression---such as the kernel stick-breaking \citep{DP08}, and the \textsc{psbp} \citep{RD11}. Indeed, as discussed in Sections \ref{sec1} and \ref{Section2}, these representations are expected to provide a comparable performance to our \textsc{lsbp} in terms of inference. However, unlike current models for Bayesian density regression, inference under the  \textsc{lsbp} is available under a broader variety of simple computational methods, thus facilitating implementation of the same model in a wider range of applications---including large $M$, $R$ and $n$ settings. Due to this, the main focus is on providing an empirical comparison of the algorithms  in Section  \ref{Section3}, while using results in \citet{DP08} as a benchmark to provide reassurance that inference under the \textsc{lsbp} is comparable to alternative representations.

\begin{figure*}[t!]
\centering
\includegraphics[width=0.93\textwidth]{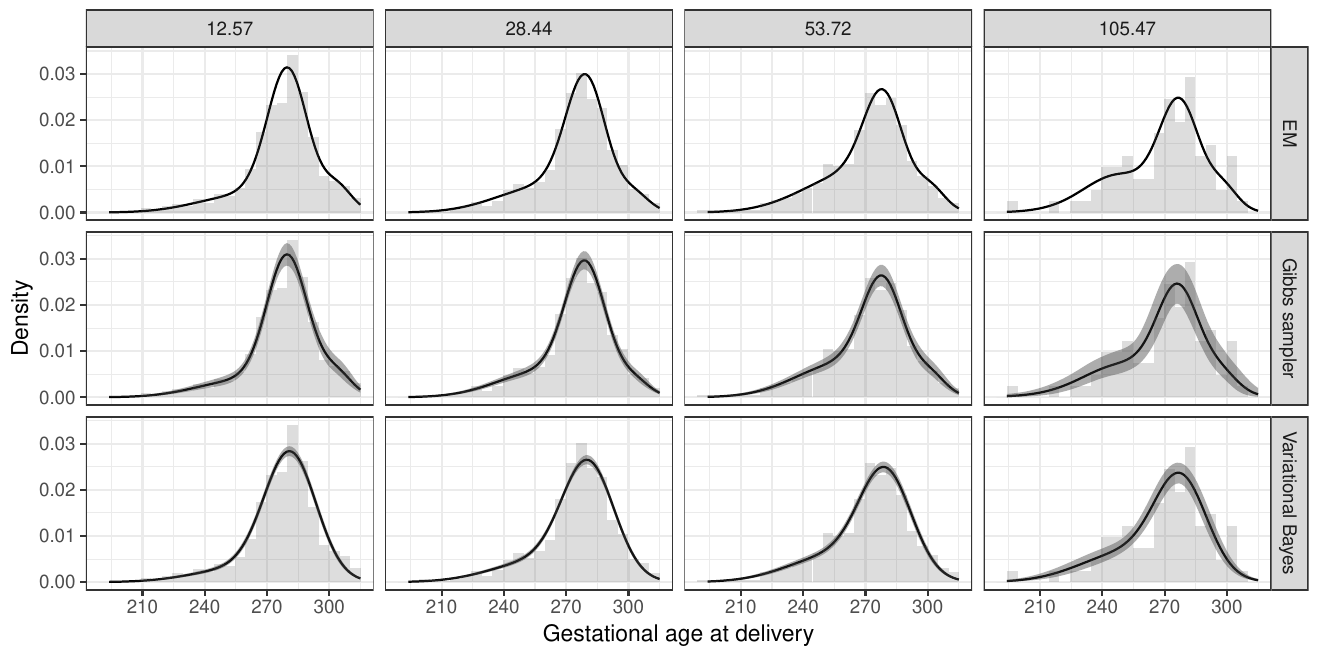}
\caption{For selected quantiles of $\texttt{DDE} \in (12.57,28.44,53.72,105.47)$, graphical representation of the posterior mean of the conditional density for $\texttt{GAD}$ given $\texttt{DDE}$, obtained from the Gibbs sampler and the \textsc{vb}, together with $0.95$ pointwise credibility intervals (shaded area). Since the \textsc{em} provides only a mode for the conditional density, we consider a graphical representation of the plug-in estimate for  $f(y \mid x)$. The histograms represent the observations of \texttt{GAD}, having $\texttt{DDE}$ in the intervals $(-\infty,20.505)$, $[20.505,41.08)$, $[41.08,79.6)$,  $[79.6,\infty)$, respectively.}
 \label{Fig::conditional}
\end{figure*}

We apply the predictor-dependent mixture of Gaussians \eqref{mix_dep_gaus} with \textsc{lsbp} \eqref{logit_stick}--\eqref{pred}, to a normalized version of the \texttt{DDE} and  \texttt{GAD} $(\bar{x}_i,\bar{y}_i)$, $i=1, \ldots, n$, and then show results for $f_x(y)$ on the original scale of the data. Consistent with previous works \citep{DP08,Canale2018}, we let $M=2$, with $\lambda_1(\bar{x}_i)=1$ and $\lambda_2(\bar{x}_i) = \bar{x}_i$, for every $i=1, \ldots,n$, and rely instead on a flexible representation for $\eta_{h}(\bar{x}_i)$ to characterize changes in the stick-breaking weights with \texttt{DDE}. In particular, each $\eta_{h}(\bar{x}_i)$ is defined via a natural cubic spline basis $\boldsymbol{\psi}(\bar{x}_i) = \{1, \psi_1(\bar{x}_i),\dots, \psi_{5}(\bar{x}_i)\}^{\intercal}$, for every $h=1, \ldots, H-1$.  Bayesian posterior inference---under the three computational methods developed in Section~\ref{Section3}---is instead performed with default hyperparameters $\boldsymbol{\mu}_{\beta}=(0,0)^{\intercal}$, $\boldsymbol{\Sigma}_{\beta}=\mbox{I}_{2\times2}$, $\boldsymbol{\mu}_{\alpha}=(0,\dots,0)^{\intercal}$, $\boldsymbol{\Sigma}_{\alpha}=\mbox{I}_{6\times 6}$ and $a_{\tau}=b_{\tau}=1$. For the total number of mixture components we consider $H=20$, and allow the shrinkage induced by the stick-breaking prior to adaptively delete redundant components not required to characterize the data. As shown in Figure \ref{Fig::conditional}, these choices allows accurate inference on $f_x(y)$.

\begin{figure*}[t!]
\centering
\includegraphics[width=0.93\textwidth]{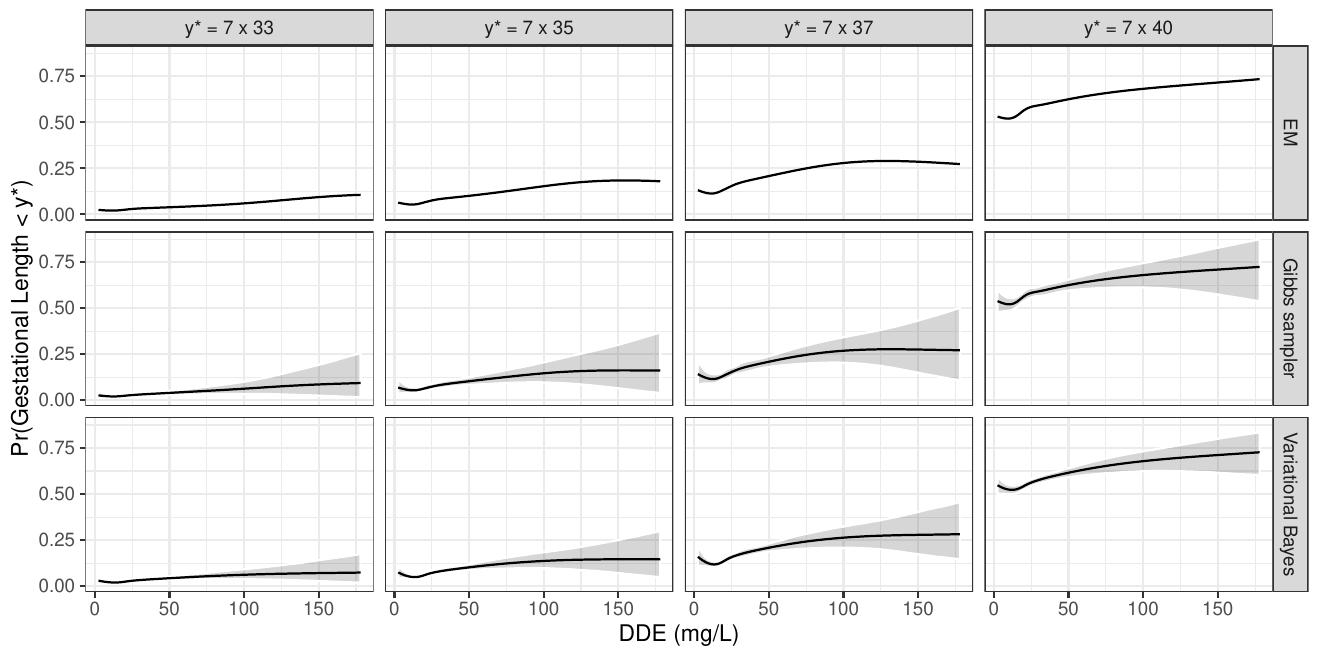}
\caption{For the Gibbs sampler and the \textsc{vb}, posterior means of four different conditional probabilities $\mbox{pr}(y < y^{*} \mid x)$---based on thresholds $y^{*} \in (7\times33,7\times35,7\times37,7\times40)$---along with $0.95$ pointwise credibility intervals (shaded area). These quantities are not available from the \textsc{em} algorithm, for which a plug-in estimate of $\mbox{pr}(y < y^{*} \mid x)$ is displayed.}
\label{Fig::posterior}
\vspace{-5pt}
\end{figure*}

In providing posterior inference under the Gibbs sampling algorithm described in Section \ref{gibbs}, we rely on $30{,}000$ iterations, after discarding the first $5{,}000$ as a burn-in, and initialize the routine from random starting values sampled from the prior. Analysis of the traceplots for the quantities discussed in Figures \ref{Fig::conditional} and \ref{Fig::posterior} showed that this choice is sufficient for good convergence. The \textsc{em} algorithm and the \textsc{vb} procedures discussed in Sections \ref{em} and \ref{vb}, respectively, are instead run until convergence to a modal solution. Since such modes could be local, we run both algorithms for different initial values, and consider the solutions having the highest log-posterior and \textsc{elbo}, respectively. We also controlled the monotonicity of the sequences for these quantities, in order to further validate the correctness of our derivations. In this study, the \textsc{em}  and the \textsc{vb} reach convergence in about $2$ and $6$ seconds, respectively, whereas the Gibbs sampler requires $5$ minutes, using a MacBook Air with a Intel Core i5.

Similarly to Figure 3 in \citet{DP08}, Figure~\ref{Fig::conditional} provides posterior inference for the conditional density $f_x(y)$ evaluated at the $0.1$, $0.6$, $0.9$, $0.99$ quantiles of \texttt{DDE}, for the three algorithms. Histograms for the \texttt{GAD}, are instead obtained by grouping the response data according to a binning of the  \texttt{DDE} with cut-offs at the central values of subsequent quantiles, so that the conditional density can be plotted alongside the corresponding histogram. Results in Figure~\ref{Fig::conditional} confirm accurate fit to the data and suggest that the left tail of the \texttt{GAD} distribution---associated with preterm deliveries---increasingly inflates as \texttt{DDE} grows. Moreover, as seen in Figure~\ref{Fig::conditional}, the three algorithms have similar results, thus providing empirical reassurance for the goodness of the proposed routines.  Although the \textsc{em} outputs a posterior mode, such an estimate matches closely the posterior mean of the Gibbs sampler, whereas the \textsc{vb} tends to over-smooth some modes of the conditional distribution.  This is likely due to the fact that the \textsc{vb} provides an approximation of the posterior, instead of the exact one. However, unlike for the \textsc{em}, this routine allows uncertainty quantification, and provides a much scalable methodology compared to the Gibbs sampler, thus representing a valid candidate in high-dimensional inference when the focus is on specific functionals of $f_x(y)$. Indeed, as shown in Figure \ref{Fig::posterior}, when the aim is to exploit $f_x(y)$ to infer conditional preterm probabilities $\mbox{pr}(y < y^{*} \mid x)=\int_{-\infty}^{y^{*}}f_x(y) \mbox{d}y$ with $y^{*} \in (7\times33,7\times35,7\times37,7\times40)$ denoting a clinical threshold, the  \textsc{vb}  provides very similar conclusions.

Prior to conclude our analysis, note that the results  in Figures \ref{Fig::conditional} and \ref{Fig::posterior} are similar to those obtained under the kernel stick-breaking prior in  \citet{DP08}. This provides empirical guarantee that the flexibility characterizing popular Bayesian nonparametric models for density regression is maintained also under  \textsc{lsbp}, which has the additional relevant benefit of facilitating computational implementation of these methodologies. Minor differences are found at extreme \texttt{DDE} exposures, but this is mainly due to the sparsity of the data in this subset of the predictor space.

\section{Discussion}
\label{Section5}
The focus of this paper has been on providing novel methods to facilitate the computational implementation of Bayesian nonparametric models for density regression in a broad range of applications. To address this goal, we have proposed an alternative reparameterization of the predictor-dependent stick-breaking weights, which relies on a set of sequential logistic regressions. This  representation has relevant connections with continuation-ratio logistic regressions and P\'olya-gamma data augmentation, thus allowing simple derivation of several algorithms of routine use in Bayesian inference. The proposed computational methods are empirically evaluated in a toxicology study, obtaining good results and reassurance that the \textsc{lsbp} maintains the same  flexibility and efficiency properties characterizing popular Bayesian nonparametric models for density regression. 

Although our dependent mixture of Gaussians provides a flexible representation, it is worth considering extensions to other kernels. For example, all our algorithms can be easily adapted to predictor-independent kernels coming from an exponential family, when  conjugate priors for their parameters are used. Similar derivations are also possible for predictor-dependent kernels within a generalized linear model representation, provided that conjugate priors for the coefficients can be found \citep[e.g.][]{CHE03}. 

\section*{Acknowledgments}

The authors are grateful to the Associate Editor and the four referees for the insightful comments and suggestions that led to a substantial improvement of the paper. This work was partially supported by grant 1R01ES027498 of the National Institutes of Environmental Health Sciences of the United States National Institutes of Health, and by the MIUR--PRIN 2017 grant 20177BRJXS.

\section*{Appendix. Proofs and additional properties of the \textsc{lsbp}}
\label{appendix}

\begin{proposition} For any fixed $ {\bf{x}} \in \mathcal{X}$, $\sum_{h=1}^{\infty} \pi_h({\bf{x}}) = 1$ almost surely, with $\pi_h({\bf{x}})$ factorized as in \eqref{logit_stick} and $\boldsymbol{\alpha}_h \sim \mbox{\normalfont N}_{R}(\boldsymbol{\mu}_{\alpha}, \boldsymbol{\Sigma}_{\alpha})$ independently for every $h\in \mathbb{N}$. Hence, the \textsc{lsbp} provides a well defined predictor-dependent random probability measure $p_{\bf x}$ at every ${\bf x} \in \mathcal{X}$.
 \label{prop1}
\end{proposition} 

\noindent{\bf Proof of Proposition~\ref{prop1}}. Recalling results in \citet{IJ01}, we have that $\sum_{h=1}^{\infty} \pi_h({\bf{x}}) = 1$ almost surely if and only if the equality $\sum_{h=1}^{\infty}\mbox{E} [\log\{1-\nu_h({\bf{x}}) \}] = -\infty$ holds. Since $\log\{1-\nu_h({\bf{x}}) \}$ is concave in $\nu_h({\bf{x}})$ for every ${\bf{x}} \in \mathcal{X}$ and $h \in \mathbb{N}$, by the Jensen inequality  $\mbox{E} [\log\{1-\nu_h({\bf{x}}) \}] \leq \log[1-\mbox{E}\{ \nu_h({\bf{x}})\} ]$. Therefore, since $\nu_h({\bf{x}}) \in (0,1)$, we have that $0<\mbox{E}\{ \nu_h({\bf{x}})\}=\mu_{\nu}({\bf{x}})<1$, thereby providing $\log\{1-\mu_{\nu}({\bf{x}})\} <0$. Leveraging these results, the proof of Proposition \ref{prop1} follows after noticing that  $\sum_{h=1}^{\infty}\mbox{E} [\log\{1-\nu_h({\bf{x}}) \}] \leq \sum_{h=1}^{\infty}\log\{1-\mu_{\nu}({\bf{x}})\}= -\infty$. 
  
 \vspace{5pt} 
\noindent{\bf Proof of Theorem~\ref{teo4}}. Adapting the proof of Theorem 1 in \citet{IJ02} to our representation we have
\begin{eqnarray*}
||f_{\bf{X}}^{(H)}({\bf{y}})  - f_{\bf{X}}^{(\infty)}({\bf{y}})||_1 \le 4 \left[ 1 - \mbox{E}\left\{\prod_{i=1}^n \sum_{h=1}^{H-1}\pi_h({\bf x}_i)\right\}\right]=4 \cdot \mbox{E}\left[ 1 -\prod_{i=1}^n \sum_{h=1}^{H-1}\pi_h({\bf x}_i)\right].
\end{eqnarray*}
Since $\sum_{h=1}^{H-1}\pi_h({\bf x}_i)\leq 1$, and $1=\prod_{i=1}^n1$, we can write $ 1 -\prod_{i=1}^n \sum_{h=1}^{H-1}\pi_h({\bf x}_i)=\prod_{i=1}^n1 -\prod_{i=1}^n \sum_{h=1}^{H-1}\pi_h({\bf x}_i) \leq \sum_{i=1}^n\{1- \sum_{h=1}^{H-1}\pi_h({\bf x}_i) \}$ \citep[][pp. 358]{BI95}.  Hence $||f_{\bf{X}}^{(H)}({\bf{y}})  - f_{\bf{X}}^{(\infty)}({\bf{y}})||_1 \leq 4[n-\sum_{i=1}^n \sum_{h=1}^{H-1}\mbox{E}\{\pi_h({\bf x}_i)\}],$ with $\sum_{h=1}^{H-1}\mbox{E}\{\pi_h({\bf x}_i)\}=\sum_{h=1}^{H-1}\mu_{\nu}({\bf x}_i) \{1-\mu_{\nu}({\bf x}_i) \}^{h-1}$ $=  1-\{1-\mu_{\nu}({\bf x}_i) \}^{H-1}$. Substituting this quantity in $4[n-\sum_{i=1}^n \sum_{h=1}^{H-1}\mbox{E}\{\pi_h({\bf x}_i)\}]$, we obtain the final bound $4\sum_{i=1}^n\{1-\mu_{\nu}({\bf x}_i) \}^{H-1}$.
\qed

%\section*{References}

\bibliography{bibliography.bib}

\end{document}